\documentclass[sigconf]{acmart}
\AtBeginDocument{%
  }
\usepackage[skins]{tcolorbox}
\usepackage{booktabs}
\usepackage{siunitx}
\usepackage{booktabs}
\usepackage{siunitx}
\usepackage{array}
\usepackage{makecell}
\usepackage{graphicx}
\usepackage{subcaption}
\usepackage{booktabs}
\usepackage{siunitx}
\usepackage{makecell}
\usepackage{caption}
\usepackage{subcaption}
\usepackage{xurl}
\usepackage{float}

\setcopyright{acmlicensed}
\copyrightyear{2026}
\acmYear{2026}
\acmDOI{XXXXXXX.XXXXXXX}

\newcommand{\rqone}{\textit{How do vulnerability reporting characteristics contribute to security risk across different vulnerability lifecycle events?}}
\newcommand{\rqtwo}{\textit{What's the exposure impact from upstream to downstream projects? }}
\newcommand{\rqthree}{\textit{Is the presence of dependency automation associated with faster downstream mitigation?}}

\newif\ifcomments
\commentsfalse    

\ifcomments
    \newcommand{\yuan}[1]{\textcolor{red}{{\it [Yuan says: #1]}}}
    \newcommand{\amit}[1]{\textcolor{blue}{{\it [Amit says: #1]}}}
    \newcommand{\mariam}[1]{\textcolor{brown}{{\it [Mariam says: #1]}}}
\else
    \newcommand{\yuan}[1]{}
    \newcommand{\amit}[1]{}
    \newcommand{\mariam}[1]{}
\fi

\begin{document}


\title{A Fine-Grained Empirical Study of Upstream Security Events and Downstream Mitigation in Maven}

\author{Kazi Amit Hasan}
\email{kaziamit.hasan@queensu.ca}
\correspondingauthor
\affiliation{%
  \institution{School of Computing, Queen's University}
  \city{Kingston}
  \state{Ontario}
  \country{Canada}
}

\author{Enrong Pan}
\email{enrong.pan@queensu.ca} 
\affiliation{%
  \institution{School of Computing, Queen's University}
  \city{Kingston}
  \state{Ontario}
  \country{Canada}
}

\author{Yuan Tian}
\email{y.tian@queensu.ca} 
\affiliation{%
  \institution{School of Computing, Queen's University}
  \city{Kingston}
  \state{Ontario}
  \country{Canada}
}

\author{Mariam Guizani}
\email{mariam.guizani@queensu.ca} 
\affiliation{%
  \institution{Department of Electrical and Computer Engineering, Queen's University}
  \city{Kingston}
  \state{Ontario}
  \country{Canada}
}

\author{Steven H. H. Ding}
\email{steven.h.ding@mcgill.ca} 
\affiliation{%
  \institution{School of Information Studies, McGill University}
  \city{Montreal}
  \state{Quebec}
  \country{Canada}
}

\renewcommand{\shortauthors}{Hasan et al.}

\begin{abstract}
Open-source vulnerability mitigation spans upstream libraries and the downstream projects that depend on them. Upstream, a vulnerability progresses through a public report when one exists, a fix, a fixed release, and NVD disclosure. Downstream, a project mitigates it by adopting the fixed release, and the timeliness of this adoption is measured relative to these events. Prior work studies the two sides separately, so it is unclear whether downstream exposure stems from delayed upstream release, delayed downstream adoption, or both. We address this gap by following 832 CVEs in the Maven ecosystem from the upstream fix to its adoption by each exposed downstream project.  We classify upstream fixes as \textit{silent} or \textit{transparent}, recover the order of report, fix, release, and NVD disclosure, and link vulnerable artifacts to their downstream dependents. We then compare this order with downstream adoption and dependabot usage. We find that delay arises on both sides. Upstream, the fixed release comes before NVD disclosure for about 82\% of CVEs, for both silent and transparent fixes. For the other 17.9\%, downstream projects have no fix to adopt for days to more than four months after disclosure. Downstream, exposure is concentrated, with the top 10\% of CVEs accounting for 72.1\% of it. Of 1,677 observed adoptions, 85.3\% happen after NVD disclosure, rising to 88.3\% for silent fixes released before disclosure. Most adoptions come through routine dependency updates. Adoptions with dependabot version updates have a median release-to-adoption delay of 10.4 days, compared with 110.6 days without dependabot. Yet version updates were configured at adoption time for only 8.5\% of total adoptions. We translate these findings into implications for upstream maintainers, downstream developers, and dependency-automation tools.

\end{abstract}


\maketitle



\section{Introduction}\label{sec:introduction}

Open-source software security is increasingly an ecosystem-level, dependency-chain problem. Modern projects rarely implement all functionality internally~\cite{alexopoulos2026cross}, relying instead on upstream libraries that are maintained, released, and disclosed independently~\cite{tang2022towards}. When an upstream library contains a vulnerability, the risk extends to every downstream project that depends on the affected artifact, and it persists there until each downstream project adopts a fixed release. Ecosystem-level vulnerability risk and mitigation therefore depend not only on upstream fixing behavior, but on the order and timing of events across the full vulnerability lifecycle: reporting, fixing, release, public disclosure, and downstream adoption. 

Prior work has examined different stages of this lifecycle, but each line of work typically holds the other stages fixed. Upstream vulnerability disclosure studies characterize how projects report, coordinate, and disclose vulnerabilities~\cite{liu2025empirical}. Security patch studies analyze when fixes become visible in public repositories and show that fix commits can appear before public disclosure~\cite{li2017large,dong2025lies}. These upstream studies do not measure how downstream projects respond. Downstream dependency migration studies measure when client projects update vulnerable libraries and quantify delays between upstream fixes, releases, public disclosure, and downstream adoption~\cite{kula2018developers,chinthanet2021lags,decan2018impact,jafari2023dependency}. Maven-specific work further studies vulnerable dependency impact by analyzing reachability, CVE lifecycles, and transitive exposure duration~\cite{wu2023understanding,yang2025tracing,przymus2025out}. Dependabot-focused studies examine automated dependency-update pull requests, including their acceptance, merge behavior, and role in security-related dependency management~\cite{alfadel2021use,mohayeji2025securing,rebatchi2024dependabot}. These downstream studies, however, typically treat upstream events as fixed reference points and measure delay from them without examining how their order varies across vulnerabilities. Because upstream fix handling and downstream adoption have been studied separately, conclusions about either side can be misattributed. An upstream-only view would consider a fix released before NVD disclosure to be well coordinated, without checking whether downstream projects adopted it before disclosure. A downstream-only view would count every day between NVD disclosure and adoption as downstream delay, without checking whether a fixed release was available at disclosure. It therefore remains unclear whether upstream event ordering gives downstream projects a window to adopt a fix before disclosure, whether downstream projects adopt within such windows, and how downstream projects adopt fixed releases, including the role of dependency automation. We address this gap by linking upstream fix transparency, event ordering, and inter-event delays to downstream exposure, adoption timing, and dependabot usage at adoption time within a single empirical framework. We apply this framework to 832 CVEs in the Maven ecosystem and their downstream dependents, and answer three research questions.


\vspace{0.1cm}
\noindent \textbf{RQ1: \rqone} We find that 19.84\% of repositories fix certain CVEs silently and others transparently. These repositories hold 58.89\% of all CVEs, so reporting behavior must be studied per CVE. Although transparent fixes make the vulnerability public earlier, reporting style is not associated with whether the fixed release comes before NVD disclosure, which holds for 81.63\% of silent fixes and 82.76\% of transparent fixes. In these cases, downstream projects can adopt the fix before the vulnerability is listed in NVD. For the remaining 17.9\% of CVEs, downstream projects stay exposed after disclosure for a few days to more than four months until the upstream release.

\vspace{0.1cm}
\noindent \textbf{RQ2: \rqtwo} In this RQ, we examine how upstream vulnerabilities propagate to downstream projects across the Maven ecosystem and how quickly downstream projects adopt fixed releases. We construct downstream exposure pairs and align observed pairs with the upstream lifecycle events identified in RQ1. Our analysis shows that exposure is highly concentrated at the ecosystem level: 832 CVEs generate 44,450 exposure pairs, and the top 10\% of CVEs account for 72.1\% of them. Among the 1,677 pairs with an observed adoption, 1,430 (85.3\%) adopt after public disclosure.


\vspace{0.1cm}
\noindent \textbf{RQ3: \rqthree}  We find that downstream projects adopt fixed releases through routine dependency updates, not in response to the vulnerability. Of the 1,677 adoption pairs, 978 (58.3\%) are linked to a PR or issue, and in a stratified sample of these, only 7.0\% mention the vulnerability. In the two most common orderings (687 of the 978 linked pairs), the PR or issue is opened a median of 105.0 and 57.5 days after NVD disclosure, and the adoption commit follows a median of 0 and 6 days later. Dependabot version updates are associated with faster adoption, with a median release-to-adoption delay of 10.4 days compared with 110.6 days without dependabot. However, although 78.2\% of downstream repositories use dependabot at some point, version updates were configured at adoption time for only 8.5\% of adoption pairs, and they do not eliminate late adoption. This paper makes three contributions:


\begin{itemize}
\item We provide an integrated analysis framework for vulnerable dependency exposure and mitigation that follows each vulnerability from the upstream project to its downstream dependents.

\item Applying this framework to the Maven ecosystem, we measure fine-grained delays, per CVE upstream and per adoption pair downstream, and show their impact on downstream exposure.

\item To the best of our knowledge, we are the first to distinguish dependabot security updates from configured version updates in this context, and analyzed their impact at the time of adoption.
\end{itemize}





\begin{figure*}[t]
    \centering
    \includegraphics[width=\linewidth]{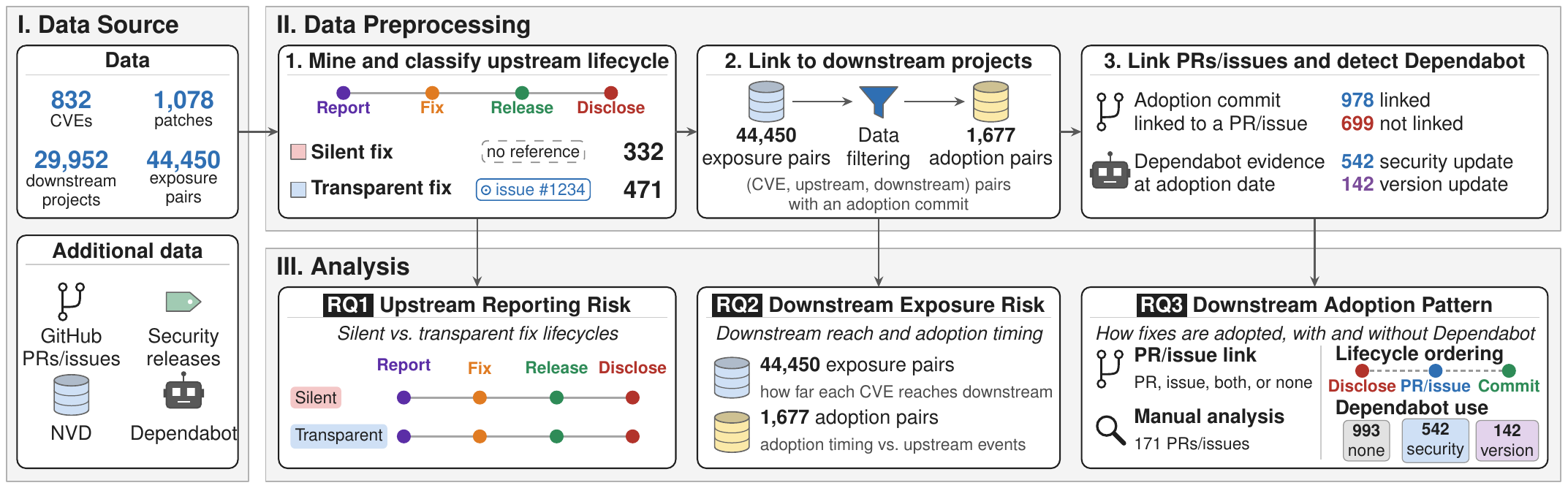}
    \caption{Overview of the study design.}
    \label{fig:overall}
\end{figure*}

\section{Terminology}\label{sec:definitions}

We define key concepts used throughout this paper:
 
\noindent \textbf{Silent fix:} A vulnerability fix that contains no vulnerability-related information (links, PRs, issues).\footnote{Example: \url{https://github.com/apache/logging-log4j2/commit/05db5f9527254632b59aed2a1d78a32c5ab74f16}}


\noindent \textbf{Transparent fix:} A vulnerability fix that contains public vulnerability-related information via GitHub PRs, issues, or any external links.\footnote{Example: \url{https://github.com/h2database/h2database/commit/d83285fd2e48fb075780ee95badee6f5a15ea7f8}}


\noindent \textbf{Vulnerability lifecycle:} The ordered sequence of a CVE's upstream report (for transparent fixes), fix, release, and NVD disclosure, extended in downstream analysis with adoption of the fixed release.


\noindent \textbf{Report date:} The creation timestamp of the vulnerability-related reporting artifact, such as a GitHub PR or issue, captured for a transparent fix. When a captured PR refers to an issue that is also captured as reporting evidence, we use the issue date.


\noindent \textbf{Fix date:} The timestamp of the upstream security fix commit that resolves the vulnerability. In case of CVEs that are fixed through multiple commits, we use the earliest commit as fix date ~\cite{li2017large}.



\noindent \textbf{Release date:} The timestamp of the first upstream release that contains all fix commits of the CVE and makes the patched version available to downstream users~\cite{chinthanet2021lags}. 



\noindent \textbf{Disclosure date:} The publication timestamp of the CVE's NVD record,\footnote{\url{https://nvd.nist.gov/vuln}} which describes the vulnerability, its affected versions, and references such as the fix commit.

\noindent \textbf{Downstream exposure pair:} A (CVE, upstream, downstream) pair that indicates the connected CVE, upstream repository, and dependent downstream artifact.

\noindent \textbf{Observed adoption pair:} A downstream exposure pair (CVE, upstream, downstream, adoption time) with an identified downstream commit that adopts the upstream fixed release.


\noindent \textbf{$T_{\text{fix}} \to T_{\text{release}}$ adoption:} An adoption event occurring after the upstream fix commit and before the upstream release. These cases occur when the downstream project updates to a \texttt{SNAPSHOT} version\footnote{\url{https://maven.apache.org/guides/getting-started/index.html\#what-is-a-snapshot-version}}, a Maven development build of the latest code that already contains the fix commit but carries no stability guarantee. 




\noindent \textbf{$T_{\text{release}} \to T_{\text{disclosure}}$ adoption:} An adoption event occurring after the upstream release and before NVD public disclosure. This interval exists only when release precedes disclosure.


\noindent \textbf{$T_{\text{disclosure}} \to$ adoption:} An adoption event occurring after NVD public disclosure.

\noindent \textbf{Exposure burden:} The number of days from the start of vulnerability-related information exposure, which varies by lifecycle pattern, to downstream adoption of the fix, computed per adoption pair and aggregated by reporting class or lifecycle pattern. For silent fixes, the exposure start depends on the order of release and NVD disclosure. When release occurs before NVD disclosure, we count from the release date because downstream projects can adopt the fixed release before disclosure. When NVD disclosure occurs before release, we count from the disclosure date because the vulnerability is already public while downstream projects remain on vulnerable versions. For transparent fixes, we count from the report date because vulnerability-related information is public through the reporting artifact, regardless of when NVD disclosure occurs.

\section{Dataset}\label{sec:dataset}
Figure \ref{fig:overall} presents an overview of the study design and illustrates how each step of the process and the resulting data artifacts contribute to each research question. We build our study on a large-scale Maven ecosystem dataset from prior work~\cite{wu2023understanding}. The baseline dataset contains 832 CVEs from 257 upstream repositories and 1,078 patches. Of these CVEs, 23.9\% are fixed through multiple commits. We mined the associated fix and release date per CVE as defined in Section~\ref{sec:definitions}. The dataset also includes 29,952 unique downstream projects that depend on these vulnerable repositories, forming 44,450 triplet relations of \(\langle\)CVE, upstream, downstream\(\rangle\). We use this dataset as the starting point and derive the RQ1, RQ2, and RQ3 analysis datasets through additional data mining, preprocessing, and filtering. 


\section{RQ1: Upstream Reporting Risk}\label{sec:rq1}
\subsection{Motivation}


Downstream projects mitigate an upstream vulnerability by adopting a fixed release. Silent fixing follows coordinated disclosure guidelines, which recommend not mentioning the vulnerability in the fix commit~\cite{cheng2025fixseeker}, but it only delays disclosure~\cite{dong2025lies}. If the fixed release precedes NVD disclosure, downstream projects can adopt it early but have no signal that it fixes a vulnerability, as they often rely on public disclosure to learn of vulnerabilities~\cite{dong2025lies}. If NVD disclosure comes first, downstream projects know about the vulnerability but have no fixed release to adopt. A transparent fix makes the vulnerability public earlier through a linked PR, issue, or advisory, but downstream projects still wait for the fixed release. Reporting style alone therefore does not determine when downstream projects can mitigate. RQ1 thus examines how upstream projects report vulnerability fixes per CVE, together with the order of the report, fix, release, and NVD disclosure and the time between them.

\subsection{Approach}

We answer RQ1 in three steps. First, we measure the prevalence of vulnerability reporting characteristics. We classify the fix of each CVE as \textit{transparent} or \textit{silent}, following the definitions in Section~\ref{sec:definitions}. This gives 471 transparent and 361 silent fixes. Second, we group upstream repositories into transparent, silent, and mixed repositories, depending on whether their CVEs have only transparent fixes, only silent fixes, or both. We compare the three groups by number of CVEs, CVE severity, and repository activity (stars, forks, closed PRs, commits, and contributors) at the fix date of each CVE. Third, we order the report, fix, release, and NVD disclosure of each CVE and compute event durations. We exclude CVEs for which one of these dates is unavailable while mining. The final dataset contains 593 CVEs (332 silent and 261 transparent). Based on this data, we compute event durations, stratified by lifecycle pattern and severity. For severity-level timing interpretation, we only interpret pattern-severity groups with at least two cases to provide meaningful analysis. We also test whether reporting style is associated with a fixed release preceding NVD disclosure using a chi-square test of independence.

\subsection{Results}

\begin{table}[]
\centering 
\caption{CVE reporting characteristics, with disclosure channel for transparent fixes.}
\label{tab:reporting-combined}
\scriptsize
\setlength{\tabcolsep}{2pt}
\begin{tabular}{l S[table-format=3.0] S[table-format=2.2]}
\toprule
\textbf{Reporting characteristic} & \textbf{CVE count} & \textbf{\%} \\
\midrule
\textbf{Transparent} & \textbf{471} & \textbf{56.61} \\
\quad GitHub PR      & 116 & 24.63 \\
\quad GitHub issues  &  71 & 15.07 \\
\quad Issue tracker  & 192 & 40.76 \\
\quad GitHub Commit  &  25 &  5.31 \\
\quad Other          &  67 & 14.22 \\
\midrule
\textbf{Silent}       & \textbf{361} & \textbf{43.39} \\
\bottomrule
\end{tabular}
\end{table}

\noindent \textbf{Finding 1.1: Transparent fixes form a majority and are reported through multiple channels.} As shown in Table~\ref{tab:reporting-combined}, among 832 CVEs from 257 repositories, 471 CVEs are transparent fixes, and 361 are silent fixes, corresponding to 56.61\% and 43.39\%, respectively. Among transparent fixes, issue trackers\footnote{https://issues.apache.org/jira/browse/AMQ-3294} are the most common identifiable reporting channel, followed by GitHub PRs and GitHub issues. The remaining 67 cases (14.22\%) include GitHub advisories or different source links such as SVN. This suggests that vulnerability reporting is not limited to GitHub-native mechanisms.

\begin{table}[t]
\centering
\caption{Repositories by reporting behavior, with associated CVEs and severity distribution}
\label{tab:repo-reporting}
\scriptsize
\setlength{\tabcolsep}{2pt}
\renewcommand{\arraystretch}{1.15}
\begin{tabular}{l r r
  S[table-format=2.0]
  S[table-format=2.0]
  S[table-format=3.0]
  S[table-format=2.0]
  S[table-format=1.0]}
\toprule
\makecell[l]{Reporting\\characteristics} & {Repository} & {CVEs} & {Crit.} & {High} & {Med.} & {Low} & {Unk.} \\
\midrule
Transparent & 110 (42.80\%) & 195 (23.44\%) &  7 & 46 & 130 & 12 & 0 \\
Silent      &  96 (37.35\%) & 147 (17.67\%) &  2 & 24 & 116 &  5 & 0 \\
Mixed       &  51 (19.84\%) & 490 (58.89\%) & 27 & 70 & 362 & 30 & 1 \\
\bottomrule
\end{tabular}
\end{table}

\begin{table}[t]
\centering 
\caption{Repository activity metrics by reporting behavior.}
\label{tab:metric-comparison}
\scriptsize
\setlength{\tabcolsep}{2pt}
\begin{tabular}{l l S[table-format=3.0] S[table-format=4.1] l}
\toprule
Metric & Reporting behavior & {n} & {Median} & {IQR} \\
\midrule
Stars &  Transparent & 190 & 160.0 & [25.25, 1328.75] \\
      &  Silent      & 146 & 321.5 & [45.25, 959.75] \\
      &  Mixed       & 489 & 721.0 & [185.00, 2499.00] \\
\midrule
Forks &  Transparent & 191 & 83.0  & [14.00, 497.00] \\
      &  Silent      & 146 & 156.0 & [27.50, 410.25] \\
      &  Mixed       & 490 & 434.5 & [100.75, 1448.50] \\
\midrule
Closed PRs & Transparent & 181 & 71.0  & [1.00, 387.00] \\
           & Silent      & 123 & 101.0 & [9.50, 397.00] \\
           & Mixed       & 460 & 238.5 & [37.75, 1262.00] \\
\midrule
Commits &  Transparent & 191 & 2973.0 & [970.00, 7290.50] \\
        &  Silent      & 146 & 2829.0 & [1459.00, 6537.00] \\
        &  Mixed       & 490 & 8425.0 & [4194.75, 15877.00] \\
\midrule
Contributors & Transparent & 191 & 35.0 & [18.00, 119.00] \\
             & Silent      & 146 & 52.0 & [23.50, 142.50] \\
             & Mixed       & 490 & 99.5 & [36.00, 202.00] \\
\bottomrule
\end{tabular}
\end{table}

\noindent \textbf{Finding 1.2: Mixed-behavior repositories concentrate vulnerability risk.} Although only 51 repositories (19.84\%) show mixed silent and transparent behavior, they account for 490 (58.89\%) of all CVEs considered. In contrast, transparent repositories represent 42.80\% of repositories but only 23.44\% of CVEs, while silent repositories represent 37.35\% of repositories and 17.67\% of CVEs. Mixed repositories also contain the largest number of severe vulnerabilities, including 27 Critical and 70 High CVEs, as shown in Table~\ref{tab:repo-reporting}. This concentration suggests that many CVEs occur in repositories that do not follow a single reporting style across vulnerabilities. In other words, a CVE-level rather than a repository-level analysis is needed to study how upstream reporting behavior affects downstream risk, because the same repository may produce both silent and transparent fixes, and each fix creates different visibility and timing conditions for downstream projects. Mixed-behavior repositories also have higher median stars, forks, closed PRs, commits, and contributors than silent and transparent repositories, suggesting that heterogeneous reporting is characteristic of high-activity upstream projects (Table~\ref{tab:metric-comparison}).


\noindent \textbf{Finding 1.3: Upstream event ordering lets downstream projects adopt a fix before NVD disclosure.} NVD disclosure publishes the vulnerability and its affected versions. If a fixed release exists by then, downstream projects can adopt it. If not, they remain exposed until the upstream project releases one. Table~\ref{tab:reporting-patterns} shows the patterns for the 593 CVEs with all lifecycle events observable. For silent fixes, S1 is the dominant pattern, accounting for 271 of 332 cases (81.6\%). In S1, the fix and the fixed release both precede NVD disclosure, which publicly announces the vulnerability and its affected versions. This ordering opens a pre-disclosure mitigation window: downstream projects that adopt the fixed release within it are never exposed while the vulnerability is publicly known. As the fix is silent, the window carries no security signal, so downstream projects would adopt within it only through routine dependency updates rather than in response to the vulnerability. In S2 (40, 12.05\%), the fix comes before NVD disclosure but the release comes after it. Unlike in S1, downstream projects learn about the vulnerability but have no fix release to adopt, although the security fix already exists. In S3 (21, 6.33\%), NVD disclosure comes before both the fix and the release. S3 is the least favorable silent pattern, because downstream projects wait for both a fix and its release after disclosure. For transparent fixes, T1 is the most common pattern (216 of 261, 82.76\%). The report comes first, followed by the fix, release, and NVD disclosure. As in S1, downstream projects can adopt the fixed release before disclosure. Unlike in S1, the report describes the vulnerability publicly before the fix, so downstream projects, and also attackers, can learn about it before disclosure. In T2 (26, 9.96\%), the report and fix come before NVD disclosure but the release comes after it. As in S2, downstream projects have no fixed release to adopt at disclosure, although the report already made the vulnerability public. In T3 (13, 4.98\%), NVD disclosure comes before the report, fix, and release, and in T4 (6, 2.30\%), the report and NVD disclosure both come before the fix and release. These are the least favorable transparent patterns, because, as in S3, downstream projects wait for both a fix and its release after disclosure. Overall, whether a fix is silent or transparent does not decide whether downstream projects have a fixed release at NVD disclosure. A fixed release exists before disclosure for 81.63\% of silent fixes (S1) and 82.76\% of transparent fixes (T1), but not for the remaining fixes (S2, S3, T2, T3, and T4). A chi-square test finds no association between reporting style and whether the fixed release precedes NVD disclosure ($\chi^2(1) = 0.13$, $p = 0.72$), and the effect size is negligible ($\phi = 0.01$).


\begin{table*}[]
\centering
\caption{Reporting patterns and associated CVE severity for silent and transparent fixes.}
\label{tab:reporting-patterns}
\scriptsize
\setlength{\tabcolsep}{6pt}
\begin{tabular}{l l l S[table-format=3.0] S[table-format=2.2] l}
\toprule
\makecell[l]{Reporting\\characteristic} & Pattern ID & Order & {Count} & {\%} & Severity distribution \\
\midrule
Silent & S1 & Fix $\to$ Release $\to$ Disclosure & 271 & 81.63 & Critical (7), High (40), Medium (211), Low (12), Unknown (1) \\
       & S2 & Fix $\to$ Disclosure $\to$ Release &  40 & 12.05 & Critical (1), High (7), Medium (30), Low (2) \\
       & S3 & Disclosure $\to$ Fix $\to$ Release &  21 &  6.33 & Critical (1), High (0), Medium (19), Low (1) \\
\cmidrule(l){2-6}
\multicolumn{2}{l}{Total} & & 332 & & \\
\midrule
Transparent & T1 & Report $\to$ Fix $\to$ Release $\to$ Disclosure & 216 & 82.76 & Critical (10), High (42), Medium (147), Low (17) \\
            & T2 & Report $\to$ Fix $\to$ Disclosure $\to$ Release &  26 &  9.96 & Critical (1), High (7), Medium (16), Low (2) \\
            & T3 & Disclosure $\to$ Report $\to$ Fix $\to$ Release &  13 &  4.98 & Critical (2), High (5), Medium (6), Low (0) \\
            & T4 & Report $\to$ Disclosure $\to$ Fix $\to$ Release &   6 &  2.30 & Critical (0), High (1), Medium (5), Low (0) \\
\cmidrule(l){2-6}
\multicolumn{2}{l}{Total} & & 261 &  & \\
\bottomrule
\end{tabular}
\end{table*}


\noindent \textbf{Finding 1.4: Upstream release delays leave downstream projects exposed after NVD disclosure.} For the 106 CVEs whose fixed release comes after NVD disclosure (Finding~1.3), the vulnerability is public while downstream projects have no fix to adopt. Their exposure during this time comes from the upstream release. Critical CVEs wait a median of 139.87 days in T3. High CVEs wait 56.93 days in S2, 53.86 days in T2, and 61.92 days in T3. Medium CVEs wait 31.92 days in S2, 70.37 days in S3, 63.63 days in T2, 35.81 days in T3, and 2.38 days in T4. Low CVEs wait 11.58 days in S2 and 11.69 days in T2. Severity therefore does not shorten the wait, as both High and Medium CVEs wait for weeks to months in the same patterns.


 \vspace{-0.1cm}
\begin{tcolorbox}[colback=white, colframe=black, boxrule=0.5pt, arc=0pt, left=6pt, right=6pt, top=4pt, bottom=4pt]
\textit{\textbf{RQ1 Summary:} Upstream projects do not follow a single reporting style, as most CVEs come from upstream repositories that fix some vulnerabilities silently and others transparently. A transparent fix usually makes the vulnerability public before NVD disclosure, while a silent fix keeps it private until then, yet reporting style is not associated with whether a fixed release is ready at disclosure (81.6\% of silent and 82.8\% of transparent fixes). Downstream exposure after disclosure therefore depends on when the upstream project releases the fix, and when the release comes after disclosure, downstream projects wait weeks to months with no fix to adopt.}
\end{tcolorbox}

\section{RQ2: Downstream Exposure Risk}\label{sec:rq2}
\subsection{Motivation}

RQ1 characterizes upstream event ordering but not its downstream effects. A fixed release protects downstream projects only if they adopt it, so the same upstream delay can lead to very different consequences across the ecosystem for two reasons. First, the number of affected downstream projects varies widely across vulnerable libraries, so a given delay translates into very different amounts of exposure. Second, downstream projects differ in how quickly they adopt fixed releases, so the same pre-disclosure window may be used by some projects and missed by others. RQ2 therefore measures downstream exposure and adoption timing across the upstream event-ordering patterns identified in RQ1.

\subsection{Approach} 


We answer RQ2 in two steps. First, we use the 44,450 downstream exposure pairs as the downstream exposure universe, which links each vulnerable upstream artifact to downstream projects that depend on the vulnerable version. This step measures downstream reach as the number of observed downstream projects per CVE in the dependency dataset. Second, we construct the observed adoption dataset. For each adoption pair, we mine the downstream adoption commit timestamp and align it with the upstream lifecycle events identified in RQ1. This step measures when downstream projects adopt fixed upstream releases, conditional on an adoption date being available. In the source dataset, adoption dates are available for a subset of downstream mitigation records. After preprocessing these records by removing duplicates, malformed downstream identifiers, incomplete upstream lifecycle cases, and missing adoption dates, the final timing dataset contains 1,677 adoption pairs from 742 downstream repositories. We do not assume that the timing subset represents all exposed downstream projects; instead, we use it only for analyses that require an adoption date. Finally, we compute exposure burden as defined in Section~\ref{sec:definitions}.



\vspace{-0.2 cm}

\begin{table}[]
\centering \setlength{\abovecaptionskip}{4pt}
\caption{Downstream exposure by CVE severity and concentration among top CVEs.}
\label{tab:exposure-combined}

\begin{subtable}{\linewidth}
\centering \setlength{\abovecaptionskip}{4pt}
\caption{Exposure by severity.} 
\label{tab:exposure-by-severity}
\scriptsize
\setlength{\tabcolsep}{2pt}
\begin{tabular}{l S[table-format=3.0] S[table-format=5.0] S[table-format=2.0] l S[table-format=4.0] S[table-format=3.1]}
\toprule
Group & {CVEs} & {Total pairs} & {Median} & {IQR} & {Max} & {\makecell{\%}} \\
\midrule
All CVEs & 832 & 44450 &  9 & [3, 29] & 4386 & 100.0 \\
\midrule
\quad Critical &  36 &  1477 &  6 & [4, 15] &  507 &  3.3 \\
\quad High     & 140 &  8823 &  9 & [4, 36] & 1605 & 19.8 \\
\quad Medium   & 608 & 28528 &  9 & [3, 29] & 4386 & 64.2 \\
\quad Low      &  47 &  5621 & 14 & [4, 42] & 4015 & 12.6 \\
\bottomrule
\end{tabular}
\end{subtable}


\begin{subtable}{\linewidth}
\centering \setlength{\abovecaptionskip}{4pt}
\caption{Concentration among top CVEs.}
\label{tab:exposure-concentration}
\scriptsize
\setlength{\tabcolsep}{2pt}
\begin{tabular}{S[table-format=2.0] S[table-format=3.0] S[table-format=2.1]}
\toprule
{Top \% of CVEs} & {CVEs} & {Share of all exposure (\%)} \\
\midrule
1  &   9 & 38.6 \\
5  &  42 & 57.5 \\
10 &  84 & 72.1 \\
25 & 208 & 89.0 \\
50 & 416 & 96.5 \\
100 & 832 & 100.0 \\
\bottomrule
\end{tabular}
\end{subtable}

\end{table}

\subsection{Results}

\noindent \textbf{Finding 2.1: Downstream reach is concentrated in a small number of vulnerabilities.} Across the full exposure universe, 832 CVEs generate 44,450 downstream exposure pairs. The median CVE reaches 9 downstream projects, but the maximum reaches 4,386 (Table \ref{tab:exposure-by-severity}). This distribution is highly concentrated, the top 1\% of CVEs account for 38.6\% of all downstream exposure, the top 10\% for 72.1\%, and the top 25\% for 89.0\% (Table \ref{tab:exposure-concentration}). The Gini coefficient is 0.813 with 95\% CI [0.749, 0.856] \cite{dorfman1979formula}. This suggests downstream exposure is not evenly distributed across upstream CVEs and small set of vulnerabilities creates most downstream reach.


\vspace{0.1cm}
\noindent \textbf{Finding 2.2: Most observed downstream adoptions occur after NVD disclosure.} To measure timing, we assign each observed adoption to one of three adoption categories based on when the adoption commit occurs relative to upstream lifecycle events. As shown in Table~\ref{tab:adoption-delay}, 1,430 of 1,677 observed adoption pairs (85.3\%) occur after NVD disclosure. Only 222 pairs (13.2\%) occur after release but before NVD disclosure, and 25 pairs (1.5\%) occur after fix but before release. Among pairs that adopt after NVD disclosure, the median adoption occurs 88.1 days after disclosure. In total, 712 pairs (42.5\%) adopt more than 90 days after disclosure.

\begin{table}[t]
  \centering \setlength{\abovecaptionskip}{4pt}
  \caption{Adoption delay by lifecycle interval and group. Medians and delay buckets are in days.}
  \label{tab:adoption-delay}
  \scriptsize
  \setlength{\tabcolsep}{1pt}
  \begin{tabular*}{\columnwidth}{@{\extracolsep{\fill}} l l r S[table-format=2.1]
    S[table-format=3.0] S[table-format=3.0] S[table-format=3.0]
    S[table-format=3.0] S[table-format=3.0]@{}}
  \toprule
  & & & & \multicolumn{5}{c}{Delay bucket (days)} \\
  \cmidrule(l){5-9}
  Group & Adoption interval & {Pairs (\%)} & {Median} & {$\leq$1}
  & {(1,7]} & {(7,30]} & {(30,90]} & {$>$90} \\
  \midrule
  All pairs & $T_{\text{fix}} \to T_{\text{release}}$        &   25 (1.5\%)   & 14.1 &   4 &   2 &  13 &   3 &   3 \\
            & $T_{\text{release}} \to T_{\text{disclosure}}$ &  222 (13.2\%)  & 10.2 &  28 &  67 &  57 &  31 &  39 \\
            & $T_{\text{disclosure}} \to$                    & 1430 (85.3\%)  & 88.1 &  68 & 173 & 201 & 276 & 712 \\
            & All                                            & 1677 (100.0\%) & {--} & 100 & 242 & 271 & 310 & 754 \\
  \midrule
  Silent & $T_{\text{fix}} \to T_{\text{release}}$        &    7 (0.6\%)   & 27.2 &   0 &   1 &   3 &   0 &   3 \\
         & $T_{\text{release}} \to T_{\text{disclosure}}$ &  111 (10.1\%)  & 13.6 &   9 &  35 &  34 &  11 &  22 \\
         & $T_{\text{disclosure}} \to$                    &  982 (89.3\%)  & 95.1 &  47 & 125 & 132 & 179 & 499 \\
         & All                                            & 1100 (100.0\%) & {--} &  56 & 161 & 169 & 190 & 524 \\
  \midrule
  Transparent & $T_{\text{fix}} \to T_{\text{release}}$        &   18 (3.1\%)   & 11.2 &   4 &   1 &  10 &   3 &   0 \\
              & $T_{\text{release}} \to T_{\text{disclosure}}$ &  111 (19.2\%)  &  8.7 &  19 &  32 &  23 &  20 &  17 \\
              & $T_{\text{disclosure}} \to$                    &  448 (77.6\%)  & 80.0 &  21 &  48 &  69 &  97 & 213 \\
              & All                                            &  577 (100.0\%) & {--} &  44 &  81 & 102 & 120 & 230 \\
  \midrule
  S1 & $T_{\text{fix}} \to T_{\text{release}}$        &    7 (0.7\%)   & 27.2 &   0 &   1 &   3 &   0 &   3 \\
     & $T_{\text{release}} \to T_{\text{disclosure}}$ &  111 (11.0\%)  & 13.6 &   9 &  35 &  34 &  11 &  22 \\
     & $T_{\text{disclosure}} \to$                    &  894 (88.3\%)  & 86.6 &  47 & 123 & 126 & 154 & 444 \\
     & All                                            & 1012 (100.0\%) & {--} &  56 & 159 & 163 & 165 & 469 \\
  \midrule
  T1 & $T_{\text{fix}} \to T_{\text{release}}$        &   11 (2.1\%)   &  8.8 &   4 &   1 &   5 &   1 &   0 \\
     & $T_{\text{release}} \to T_{\text{disclosure}}$ &  111 (21.3\%)  &  8.7 &  19 &  32 &  23 &  20 &  17 \\
     & $T_{\text{disclosure}} \to$                    &  399 (76.6\%)  & 68.4 &  21 &  47 &  66 &  84 & 181 \\
     & All                                            &  521 (100.0\%) & {--} &  44 &  80 &  94 & 105 & 198 \\
  \bottomrule
  \end{tabular*}
\end{table}

\vspace{0.1cm}
\noindent \textbf{Finding 2.3: Upstream pattern, S1 provides the strongest silent adoption opportunity, but downstream projects usually miss it.}
RQ1 identifies S1 as the least exposure prone silent pattern because fix and release occur before NVD disclosure and no public report appears before the fix. Of the 271 S1 CVEs, 95 have observed downstream adoption, yielding 1,012 pairs. Despite this favorable lifecycle ordering, 894 of the 1,012 S1 pairs (88.3\%) adopt after disclosure, as shown in Table~\ref{tab:adoption-delay}. For these 894 pairs, the median delay after disclosure is 86.6 days. Only 111 S1 pairs (11.0\%) adopt between release and NVD disclosure, with a median delay of 13.6 days after release.

\vspace{0.1cm}
\noindent \textbf{Finding 2.4: 20.4\% downstream projects adopt fix within a week, while most take more than a month.} Table \ref{tab:adoption-delay} shows a visible fast adoption subpopulation. Across all 1,677 observed adoption pairs, 100 pairs (6.0\%) adopt within one day, and another 242 pairs (14.4\%) adopt within 1-7 days. Combined, 342 pairs (20.4\%) adopt within the first week. The pattern is also visible in the silent and transparent groups. Transparent pairs show slightly more same-day adoption than silent pairs (7.6\% vs. 5.1\%). In T1 specifically, 44 pairs (8.4\%) adopt within one day, compared with 56 (5.5\%) in S1. Within $T_{\text{release}} \to T_{\text{disclosure}}$, T1 also has more first-week adoption than S1: 51 (45.9\%) vs. 44 (39.6\%). These very short delays are unlikely to reflect manual dependency review. One explanation is that some downstream projects run automated update workflows that surface new releases without human monitoring, a hypothesis we test directly in RQ3.

\begin{table}[]
\centering \setlength{\abovecaptionskip}{4pt}
\caption{Downstream exposure burden (in days) by reporting class and pattern.}
\label{tab:burden-by-pattern}
\scriptsize
\setlength{\tabcolsep}{3pt}
\begin{tabular}{l l S[table-format=4.0] S[table-format=6.1] S[table-format=2.1]}
\toprule
Patterns & Burden from & {Pairs} & {Burden} & {\% total} \\
\midrule
\textbf{Silent} & & 1100 & 355295.6 & 67.4 \\
\quad S1 & release    & 1012 & 332689.8 & 63.1 \\
\quad S2 & disclosure &   68 &  18051.0 &  3.4 \\
\quad S3 & disclosure &   20 &   4554.8 &  0.9 \\
\midrule
\textbf{Transparent} & & 577 & 171824.0 & 32.6 \\
\quad T1 & report & 521 & 148849.1 & 28.2 \\
\quad T2 & report &  34 &  15902.7 &  3.0 \\
\quad T3 & report &  21 &   6983.3 &  1.3 \\
\quad T4 & report &   1 &     88.9 &  0.0 \\
\bottomrule
\end{tabular}
\end{table}

\vspace{0.1cm}
\noindent \textbf{Finding 2.5: Most exposure burden comes from S1, the most common pattern, despite its pre-disclosure window.} The observed dataset accumulates 527,119.6 downstream days (1,444.2 project-years) of exposure (Table~\ref{tab:burden-by-pattern}). The 527,119 days come from 1,677 observed pairs, 3.8\% of the 44,450 exposure universe. And those are precisely the pairs where adoption eventually happened and was observable. Silent patterns account for 355,295.6 downstream-days (67.4\%), and transparent patterns account for 171,824.0 downstream-days (32.6\%). S1 contributes the largest burden, 63.1\% of the total burden (Table~\ref{tab:burden-by-pattern}). T1 is the largest transparent contributor with 148,849.1 downstream-days (28.2\%). S1's aggregate burden reflects its prevalence in the observed adoption dataset. Although S1 is the least exposure-prone silent pattern in RQ1, its downstream adoption opportunity is rarely utilized. Thus, downstream impact is not determined by lifecycle ordering alone; it depends on both pattern prevalence and adoption timing.

\vspace{-0.1cm}
\begin{tcolorbox}[colback=white, colframe=black, boxrule=0.5pt, arc=0pt, left=6pt, right=6pt, top=4pt, bottom=4pt]
\textit{\textbf{RQ2 Summary:} Downstream exposure is concentrated in a small number of vulnerabilities, and most observed downstream adoptions occur after public disclosure. Even S1, the strongest silent adoption opportunity, is usually not utilized by downstream projects. Overall, upstream to downstream impact depends on both downstream reach and adoption timing.}
\end{tcolorbox}

\section{RQ3: Downstream Adoption Patterns}\label{sec:rq3}

\subsection{Motivation}
RQ2 shows when downstream projects adopt fixed upstream releases, but an adoption commit alone does not show how the update reached the downstream repository: through a pull request (PR), through an issue, or through a direct commit. Tracing this path and its timing relative to NVD disclosure reveals where adoption delay arises, either before a PR or issue is opened or between its opening and the adoption commit. RQ2 also identifies rapid adoptions, which raises the question of whether dependency automation helps projects update sooner. RQ3 therefore examines (i) the ordering of NVD disclosure, PR/issue creation, and the adoption commit, and (ii) whether dependency automation, specifically dependabot and its features, is associated with shorter release-to-adoption delay.

\subsection{Approach}

We begin with the 1,677 observed adoption pairs from RQ2. For each adoption commit, we use the GitHub API to retrieve the PR that contains the commit and the issues linked to that PR or referenced in the commit message. As shown in Table~\ref{tab:rq3-records}, 978 pairs (58.3\%) are linked to at least one PR or issue, and 699 (41.7\%) are linked to neither. For each linked pair, we take the NVD disclosure date, the creation date of the linked PR/issue, and the adoption commit date. When a pair has several linked PRs/issues, we use the earliest creation date, because it is the first point at which the update is visible in the downstream repository. We then order the three dates and group pairs with the same order into lifecycle patterns, treating events on the same day as one step. For each pattern, we compute the median delay between consecutive events.

\begin{table}[t]
  \centering
  \caption{Connected downstream records for the 1,677 adoption pairs.}
  \label{tab:rq3-records}
  \scriptsize
  \begin{tabular}{l r}
  \toprule
  Connected records & Adoption pairs \\
  \midrule
  PR only & 851 \\
  Issue only & 71 \\
  PR and Issue & 56 \\
  Neither found & 699 \\
  \midrule
  Total & 1,677 \\
  \bottomrule
  \end{tabular}
\end{table}

To examine whether dependency automation is associated with shorter adoption delay across all 1,677 pairs, we focus on dependabot because of its wide use in dependency management~\cite{rebatchi2024dependabot}. For each of the 742 downstream repositories, we mine dependabot evidence from its GitHub history since 2017, when dependabot was introduced. Dependabot provides several dependency management mechanisms. Dependabot alerts notify maintainers about vulnerable dependencies, dependabot security updates open PRs for dependencies with known vulnerabilities, and dependabot version updates open PRs for new dependency versions based on a repository configuration file.\footnote{\url{https://docs.github.com/en/code-security/tutorials/secure-your-dependencies/dependabot-quickstart}} We do not measure alerts, because alert notifications and repository security settings are not consistently visible in public repository history. We operationalize dependabot usage at the adoption date using two observable signals. First, we identify the first dependabot-generated PR in the downstream repository. If such a pull request exists on or before the downstream adoption date, but no dependabot configuration file exists by that date, we classify the pair as having \textit{security-update evidence}. This captures cases where dependabot is visible through generated pull requests, but version-update configuration is not yet present. Second, we identify the first creation date of \textit{.github/dependabot.yml} or \textit{.github/dependabot.yaml}. If this configuration file exists on or before the downstream adoption date, we classify the pair as \textit{version-update enabled}, because dependabot version updates require this file. This adoption-time classification is necessary because the same downstream repository can appear in multiple adoption pairs at different dates. A repository may have no dependabot evidence for an earlier adoption event but show dependabot evidence for a later event. Therefore, we classify dependabot status separately for each adoption pair, rather than assigning a single repository-level label. We then compare release-to-adoption delay across three groups using the Mann-Whitney U test \cite{mcknight2010mann}.

\vspace{-0.2 cm}

\subsection{Results}

\noindent\textbf{Finding 3.1: 41.7\% of downstream adoption commits have no linked GitHub PR or issue.} Among the 1,677 adoption pairs, 851 adoption commits are linked to a PR only, 71 to an issue only, and 56 to both (Table~\ref{tab:rq3-records}). The remaining 699 pairs (41.7\%) have no linked GitHub PR or issue. At adoption, \textit{SECURITY.md}\footnote{https://docs.github.com/en/code-security/how-tos/report-and-fix-vulnerabilities/configure-vulnerability-reporting/add-security-policy} was present in 5.4\% of PR-only pairs, 2.8\% of issue-only pairs, 17.9\% of pairs linked to both, and 1.6\% of resolved pairs with neither link. Thus, many adoptions leave no public trace, and few projects document a security policy. For the 978 linked pairs, the PR/issue shows when the update first appears, letting us locate the delay.

\noindent\textbf{Finding 3.2: Most of the delay occurs before the PR/issue is opened.} Ordering the three dates of the 978 linked pairs yields 11 lifecycle patterns (D1--D11). Finding~2.2 shows that 85.3\% of adoption pairs occur after NVD disclosure. To locate where this delay arises, we divide the patterns into three groups (G1--G3) by when disclosure occurs (Table~\ref{tab:lifecycle-patterns}). In G1, disclosure occurs on or before both the PR/issue and the adoption commit. In G2, the PR/issue precedes disclosure but the adoption commit follows it. In G3, the adoption commit precedes disclosure.
 
In G1, which holds 84.3\% of linked pairs, the delay falls before the PR/issue is opened. In D1 and D2 (687 pairs), the PR/issue is opened a median of 105.0 and 57.5 days after disclosure, and the adoption commit follows a median of 0 and 6 days later. D4 is the slowest pattern, with the adoption commit a median of 254.0 days after disclosure. In G2, the delay instead falls after the PR/issue is opened. In D6, the PR/issue is opened a median of 74.0 days before disclosure, yet the adoption commit comes a median of 45.0 days after disclosure. In G3, the adoption commit follows the PR/issue within days, a median of 0 and 3 days later in D3 and D7.


\noindent\textbf{Finding 3.3: Adoption follows the project's routine update process rather than vulnerability disclosure.} To understand why adoption is delayed, we manually inspect a sample of linked PRs/issues from each group using stratified random sampling~\cite{cohen2025stratified}. Since we report results per group, we determine each group's sample size separately for a 95\% confidence level and a 10\% margin of error, yielding 87, 31, and 53 pairs for G1, G2, and G3, respectively (171 pairs in total). Within each group, we use proportional allocation~\cite{chen2001proportional}, in which each pattern contributes pairs in proportion to its size. For each sampled pair, we check who opened the PR/issue (a bot or a developer) and whether its title or description mentions the vulnerability (e.g., its CVE ID). Together, these two checks show whether the update was initiated by automation or by a developer, and whether it was related to security or to routine dependency maintenance. The sampled PRs/issues explain the timing in Finding~3.2 through three patterns. \textit{(1) In all groups, the PR/issue is a routine update.} Only 7 of 87 (8.0\%), 2 of 31 (6.5\%), and 3 of 53 (5.7\%) sampled PRs/issues in G1, G2, and G3 mention the vulnerability, and only 5 of the 102 opened by developers do. This holds even in G1, where the vulnerability is public long before the PR/issue is opened. This suggests that the fixed release mostly reaches the project through a regular dependency update, not because developers acted on the published vulnerability. \textit{(2) An early PR/issue does not guarantee an early fix.} In both G2 and G3, the PR/issue is typically opened before disclosure, but only in G3 is it merged within days. In G2, it stays open until after disclosure, and since it rarely mentions the vulnerability (pattern 1), disclosure does not appear to speed it up. \textit{(3) Routine updates opened by bots are adopted much sooner.} Bots, mostly dependabot (63 of 69), opened 35 of 87, 10 of 31, and 24 of 53 sampled PRs/issues in G1, G2, and G3, respectively. In every group, these are adopted sooner after the fixed release than developer-opened ones (9.0 vs.\ 89.0 days overall). Bots do so without flagging the vulnerability (only 7 of 69 mention it).

\begin{table}[t]
  \centering \setlength{\abovecaptionskip}{3pt}
  \caption{Delay between consecutive events by downstream project pairs. Patterns are grouped (G1--G3) by when NVD disclosure occurs relative to the PR/issue and the adoption commit. Medians and delay buckets are in days.}
  \label{tab:lifecycle-patterns}
  \scriptsize
  \setlength{\tabcolsep}{1pt}
  \begin{tabular*}{\columnwidth}{@{\extracolsep{\fill}} l l c S[table-format=3.1]
    S[table-format=3.0] S[table-format=3.0] S[table-format=3.0]
    S[table-format=3.0] S[table-format=3.0]@{}}
  \toprule
  & & & & \multicolumn{5}{c}{Delay bucket (days)} \\
  \cmidrule(l){5-9}
  Pattern & Interval & {Pairs (\%)} & {Median} & {$\leq$1}
  & {(1,7]} & {(7,30]} & {(30,90]} & {$>$90} \\
  \midrule
  \multicolumn{9}{@{}l}{\textbf{G1:} NVD disclosure first (824 pairs, 84.3\%)} \\
  \cmidrule{1-9}
  D1  & NVD $\to$ PR/Issue    & 445       & 105.0 &   3 &  87 &  43 &  77 & 235 \\
      & PR/Issue $\to$ Commit & (45.5\%)  &   0.0 & 445 &   0 &   0 &   0 &   0 \\ \addlinespace[2pt]
  D2  & NVD $\to$ PR/Issue    & 242       &  57.5 &   1 &  42 &  45 &  50 & 104 \\
      & PR/Issue $\to$ Commit & (24.7\%)  &   6.0 &  63 &  79 &  41 &  35 &  24 \\ \addlinespace[2pt]
  D4  & NVD $\to$ Commit      &  66       & 254.0 &   1 &   3 &   6 &  14 &  42 \\
      & Commit $\to$ PR/Issue & (6.7\%)   &  27.0 &   9 &  10 &  16 &  16 &  15 \\ \addlinespace[2pt]
  D5  & NVD $\to$ PR/Issue    &  48       &   0.0 &  48 &   0 &   0 &   0 &   0 \\
      & PR/Issue $\to$ Commit & (4.9\%)   &   0.0 &  48 &   0 &   0 &   0 &   0 \\ \addlinespace[2pt]
  D8  & NVD $\to$ PR/Issue    &  23       &   0.0 &  23 &   0 &   0 &   0 &   0 \\
      & PR/Issue $\to$ Commit & (2.4\%)   &  11.0 &   5 &   4 &   4 &   0 &  10 \\
  \addlinespace[5pt]
  \multicolumn{9}{@{}l}{\textbf{G2:} NVD disclosure between PR/issue and commit (43 pairs, 4.4\%)} \\
  \cmidrule{1-9}
  D6  & PR/Issue $\to$ NVD    &  30       &  74.0 &   3 &   5 &   3 &   8 &  11 \\
      & NVD $\to$ Commit      & (3.1\%)   &  45.0 &   0 &   7 &   4 &   8 &  11 \\ \addlinespace[2pt]
  D9  & PR/Issue $\to$ NVD    &  13       &   8.0 &   1 &   4 &   8 &   0 &   0 \\
      & NVD $\to$ Commit      & (1.3\%)   &   0.0 &  13 &   0 &   0 &   0 &   0 \\
  \addlinespace[5pt]
  \multicolumn{9}{@{}l}{\textbf{G3:} NVD disclosure last (111 pairs, 11.3\%)} \\
  \cmidrule{1-9}
  D3  & PR/Issue $\to$ Commit &  74       &   0.0 &  74 &   0 &   0 &   0 &   0 \\
      & Commit $\to$ NVD      & (7.6\%)   &  10.0 &   5 &  28 &  18 &  21 &   2 \\ \addlinespace[2pt]
  D7  & PR/Issue $\to$ Commit &  29       &   3.0 &  11 &  10 &   2 &   4 &   2 \\
      & Commit $\to$ NVD      & (3.0\%)   &  21.0 &   1 &  10 &   4 &  13 &   1 \\ \addlinespace[2pt]
  D10 & Commit $\to$ PR/Issue &   5       &   4.0 &   0 &   3 &   0 &   1 &   1 \\
      & PR/Issue $\to$ NVD    & (0.5\%)   &  56.0 &   0 &   1 &   0 &   3 &   1 \\ \addlinespace[2pt]
  D11 & Commit $\to$ NVD      &   3       &   6.0 &   1 &   1 &   1 &   0 &   0 \\
      & NVD $\to$ PR/Issue    & (0.3\%)   &  55.0 &   0 &   0 &   1 &   2 &   0 \\
  \midrule
  \multicolumn{2}{@{}l}{Total} & 978 (100.0\%) & & & & & & \\
      
  \bottomrule
  \end{tabular*}
\end{table}

\noindent \textbf{Finding 3.4: Dependabot is common at the repository level.} The 1,677 adoption pairs come from 742 downstream repositories, and 580 of these repositories (78.2\%) use dependabot at some point in their history. A repository can adopt different fixed releases on different dates, so we check whether dependabot was already in use at each adoption date. Only 684 pairs (40.8\%) have dependabot evidence by the adoption date (Table~\ref{tab:rq3}), and version updates are configured for only 142 of them (8.5\% of all pairs). Repository-level usage therefore overstates how often dependabot is in place when a downstream project adopts a fixed release.

\begin{table}[t]
\centering \setlength{\abovecaptionskip}{4pt}
\caption{Distribution of dependabot and its feature usage.}
\label{tab:rq3}
\scriptsize
\setlength{\tabcolsep}{2pt}
\renewcommand{\arraystretch}{1.15}
\begin{tabular}{l r S[table-format=3.1] r r}
\toprule
\makecell{Adopted\\feature} & \makecell{Pairs} & {\makecell{Median\\delay (days)}} & \makecell{$>$30 days\\late} & \makecell{$>$90 days\\late} \\
\midrule
No dependabot   & 993 (59.2\%)      & 110.6 & 711 (71.6\%)     & 525 (52.9\%) \\
Security update & 542 (32.3\%)      &  73.2 & 346 (63.8\%)     & 254 (46.9\%) \\
Version update  & 142 (8.5\%)       &  10.4 &  60 (42.3\%)     &  44 (31.0\%) \\
\midrule
Total           & 1{,}677 (100.0\%) &  85.6 & 1{,}117 (66.6\%) & 823 (49.1\%) \\
\bottomrule
\end{tabular}
\end{table}

\vspace{0.1cm}
\noindent \textbf{Finding 3.5: At adoption time, downstream projects use dependabot’s security update feature 3.8x more than others.}  Among the projects that are using dependabot at adoption time, we find 79\% of pairs are using dependabot’s security update feature and remaining 21\% are using version update feature (Table \ref{tab:rq3}). We also find that, among the 580 dependabot adopted downstream repositories, 147 repositories adopted security update feature first, and later adopted version update feature of dependabot. But the median transition time is 585.0 days. This suggests repositories are using dependabot and its features.

\vspace{0.1cm}
\noindent \textbf{Finding 3.6: Dependabot is associated with shorter release- to-adoption delay.} We compare dependabot groups using release-to-adoption delay, measured from the upstream release date to the downstream adoption date. Downstream adoption pairs with no dependabot evidence have a median delay of 110.6 days. Pairs with security-update evidence have a median delay of 73.2 days, while pairs with version-update configuration have a median delay of 10.4 days. The binary comparison between no dependabot and any dependabot evidence also shows a lower median delay for dependabot-at-adoption cases (110.6 days vs. 66.8 days). A Mann-Whitney U test indicates that these distributions differ significantly (p$<$0.05)~\cite{mcknight2010mann}. Version-update configuration is also associated with significantly shorter delay than security-update evidence; a one-sided Mann-Whitney U test shows that the two distributions differ significantly (p$<$0.05)~\cite{mcknight2010mann}. These results indicate that dependabot is associated with faster fixed-release adoption, but the magnitude differs by feature. Version-update configuration has the shortest median delay, suggesting that explicit dependency-update configuration is associated with more timely downstream mitigation than security-update evidence alone. A plausible explanation follows from Finding~3.2. Version updates open a PR once a new upstream version is released, whereas security updates open a PR only after the vulnerability is published as a security advisory. Version updates therefore shorten the time before the PR/issue is opened, which is where most of the delay occurs.

\vspace{0.1cm}

\noindent \textbf{Finding 3.7: Dependabot does not eliminate late adoption.}  Even when dependabot exists by the adoption date, many downstream projects still adopt upstream security releases very late. For adoption pairs with no dependabot, 71.6\% were more than 30 days late, and 52.9\% were more than 90 days late (Table~\ref{tab:rq3}). For security-update pairs, 63.8\% were more than 30 days late, and 46.9\% were more than 90 days late. For version-update-enabled pairs, 42.3\% were more than 30 days late, and 31.0\% were more than 90 days late. Dependabot reduces delay but does not remove downstream exposure. This is consistent with pattern (2) in Finding~3.3. In G2, dependabot opened 9 of the 31 sampled PRs/issues before disclosure, yet the fix came only after disclosure. Automation shortens the time until a PR is opened, but adoption still depends on developers merging it.

\vspace{-0.1cm}
\begin{tcolorbox}[colback=white, colframe=black, boxrule=0.5pt, arc=0pt, left=6pt, right=6pt, top=4pt, bottom=4pt]
\textit{\textbf{RQ3 Summary:} 
Downstream projects adopt fixes through routine dependency updates rather than in response to vulnerability disclosure. Dependency automation, especially dependabot version updates, shortens this wait by opening PRs for new releases. However, few projects utilize its features effectively.}
\end{tcolorbox}


\section{Implications} \label{sec:disc}

\textbf{\textit{Hosting platforms such as GitHub could provide an encrypted channel for sharing security fixes privately before disclosure.}} For 271 of 332 silent fixes, the fixed release precedes NVD disclosure (Finding~1.3). However, because the release does not indicate that it fixes a vulnerability, adoption occurs only after disclosure in 894 of the 1,012 S1 adoption pairs (88.3\%) (Finding~2.3). Coordinated disclosure, which most practitioners support~\cite{liu2025empirical}, discourages announcing such fixes before disclosure~\cite{cheng2025fixseeker}. As GitHub hosts both upstream and downstream projects, it is in a position to connect their maintainers through encrypted messages accessible only to verified maintainers. Through this channel, upstream maintainers could share the fix commit once it is made, enabling dependents with urgent needs to adopt it through a \texttt{SNAPSHOT}\footnote{\url{https://maven.apache.org/guides/getting-started/index.html\#what-is-a-snapshot-version}} build, as 25 adoption pairs did (Finding~2.2), although such builds may be unstable. Once the fixed release is published, upstream maintainers would share the fixed version and the expected disclosure date, so that dependents can adopt the release before the vulnerability becomes public. Downstream maintainers could, in turn, report their adoption, providing upstream maintainers with an estimate of how many dependents remain exposed. All participants would be required to keep these messages confidential until disclosure~\cite{householder2017cert}, consistent with the existing practice of sharing fixes among trusted maintainers~\cite{ramsauer2020sound}.

\textbf{\textit{Downstream maintainers should treat dependency automation as an early protection measure.}} RQ3 shows that, at the repository level, 580 of 742 repositories (78.2\%) adopted dependabot at some point, but only 684 of 1,677 adoption pairs (40.8\%) had dependabot evidence by the adoption date. This gap means many repositories eventually adopt automation, but not early enough to assist with the observed vulnerable dependency adoption events. Downstream maintainers cannot predict a specific future vulnerability. Therefore, dependency automation should be enabled before vulnerability-specific events occur, so update pull requests can be generated when fixed releases become available.

\textbf{\textit{The choice of dependabot feature matters.}} Adoption pairs with no dependabot have a median release-to-adoption delay of 110.6 days. Security update enabled pairs reduce this median to 73.2 days, while version update enabled pairs reduce it to 10.4 days. This suggests that scheduled version update workflows provide stronger mitigation support than other features. Downstream projects should therefore prioritize explicit version update configuration.

\textbf{\textit{Dependency management tools should facilitate the adoption of fixed releases, not only the detection of vulnerabilities.}} An automated update PR mitigates a vulnerability only after the fixed release is adopted, yet even with version updates enabled, 31.0\% of pairs are adopted more than 90 days after release (Finding~3.7). Our findings suggest three improvements. First, once a vulnerability is disclosed, tools could flag already-open routine update PRs that fix it (Finding~3.3). Second, tools should report whether the vulnerable code is reachable from the project~\cite{ponta2018beyond}, helping maintainers prioritize the update. Third, since breaking changes deter developers from upgrading~\cite{ochoa2022breaking}, tools could explain build failures caused by the update~\cite{reyes2024breaking}, helping maintainers merge it sooner.

\textbf{\textit{Vulnerable dependency mitigation should be studied as a linked upstream to downstream process.}} Delay arises on both sides. On the upstream side, the fixed release comes after NVD disclosure for 17.9\% CVEs, so downstream projects have no fix to adopt until the upstream release (Findings~1.3 and 1.4). On the downstream side, 1,430 of 1,677 adoptions (85.3\%) happen after NVD disclosure, including 88.3\% S1 adoptions pairs whose release came before disclosure (Findings~2.2 and 2.3). A study that observes only one side can therefore attribute delay to the wrong side. A downstream-only study that measures delay from NVD disclosure would count the upstream wait as downstream delay. An upstream-only study would consider a release before disclosure sufficient, overlooking that downstream projects still adopt it after disclosure. Future studies should therefore link upstream and downstream events to measure each delay where it arises.

\textbf{\textit{Automation studies should distinguish dependency management modes at the adoption event.}} Finding 3.3 and 3.6 show that treating dependabot as a binary variable hides important differences. In our data, no-dependabot pairs have a median release-to-adoption delay of 110.6 days, security-update pairs have a median delay of 73.2 days, and version-update pairs have a median delay of 10.4 days. Future studies and tools should distinguish alerting, security-update pull requests, and scheduled version-update configurations rather than collapsing them into a single bot-usage measure.

\textbf{\textit{Studies that measure mitigation through bot PR outcomes miss a large share of adoptions and observe mainly the fastest ones.}} Prior studies of dependency automation measure mitigation through the outcomes of bot PRs, such as their merge rate, close rate, and time to merge~\cite{alfadel2021use, rebatchi2024dependabot, he2023automating, mohayeji2023investigating}. Mohayeji et al.~\cite{mohayeji2025securing} report that developers often fix vulnerabilities manually when they do not merge a dependabot security update. Our results show that, in Maven, many adoptions bypass PRs and issues entirely. Of the 1,677 adoption pairs, 699 (41.7\%) have no linked PR or issue on GitHub, so a PR-based measure observes at most 907 adoptions (54.1\%) (Finding~3.1). A bot-PR-based measure observes even fewer, because bots opened only 69 of the 171 sampled PRs/issues (40.4\%) (Finding~3.3). These bot-opened updates are also adopted much sooner than developer-opened ones (9.0 vs.\ 89.0 days), so a bot-PR-based measure observes mainly the fastest adoptions and overstates how quickly downstream projects mitigate. Mitigation studies should therefore measure adoption from the downstream adoption commit and treat bot PRs as one of several adoption paths.

\section{Threats to Validity} \label{sec:threats}
Our classification relies on publicly observable evidence, but handling may still have occurred through private reports, embargoed communication, or mailing lists. Lifecycle ordering depends on mined timestamps, which are subject to timezone differences, delayed release tagging, rewritten history, and incorrect CVE-to-commit links. We retain only cases with complete lifecycle dates to reduce this risk. Repositories with more CVEs have more opportunities to exhibit both fix types, so we interpret \textit{mixed} as observed reporting heterogeneity rather than evidence that it causes higher vulnerability counts.  We use the NVD publication date as the disclosure date, although a vulnerability can become public earlier through other sources. We observe adoption for only 1,677 of the 44,450 exposure pairs, so our adoption timing results describe only projects whose adoption we could observe. We infer dependabot usage from dependabot-generated PRs and configuration files, which do not capture dependabot alerts or other dependency automation tools. Finally, our results cover CVE-level vulnerabilities in open-source Maven repositories and may not generalize to vulnerabilities without CVEs, private or closed-source projects, or ecosystems with different advisory, release, and dependency-management practices.

\section{Related Work}\label{sec:related_work}
\noindent\textbf{Vulnerability management:} Several empirical studies have examined vulnerability management in open-source software ecosystems. Alfadel et al.~\cite{alfadel2023empirical} analyzed security vulnerabilities in PyPI and showed that many vulnerabilities are remediated after public disclosure. Decan et al.~\cite{decan2018impact} studied vulnerability propagation in the NPM ecosystem and showed how vulnerable packages can affect a large number of downstream dependents. Zhang et al.~\cite{zhang2023mitigating} studied persistent vulnerabilities in Maven and proposed Ranger to identify secure version ranges for known vulnerabilities. Imtiaz et al.~\cite{imtiaz2022open} analyzed temporal gaps between security fixes and releases, showing that delays in patch deployment can affect vulnerability mitigation. These studies examine individual parts of the mitigation process, but do not connect upstream reporting behavior, fix timing, release timing, disclosure timing, downstream dependency reach, and downstream fixed-release adoption in one event-level analysis. Our work addresses this gap by analyzing vulnerable dependency mitigation as an upstream-to-downstream lifecycle.

\noindent\textbf{Dependabot:} Several empirical studies have examined dependabot in vulnerability management. Alfadel et al.~\cite{alfadel2021use} manually investigated why dependabot PRs are not accepted. Rebatchi et al.~\cite{rebatchi2024dependabot} analyzed dependabot's role in security-related PRs and found that it accounts for about 65\% of dependency management activity, owing to its efficiency, adaptability, ease of access, and support availability. Mohayeji et al.~\cite{mohayeji2023investigating, mohayeji2025securing} analyzed how long dependabot PRs take from submission to merge and the factors that affect this duration. These studies focus on dependabot PRs, merge rates, and general usage, but do not examine which dependabot features are in place when a project adopts a fixed release. Our work classifies dependabot features at each adoption date and relates them to release-to-adoption delay.

\section{Conclusion}\label{sec:conclusion}

Analyzing upstream lifecycle patterns, timing, and downstream adoption together shows that neither alone explains downstream exposure: coordinated upstream releases create adoption windows that downstream projects largely do not use, and automation shortens delay without closing it. Mitigation is a joint property of release timing, downstream responsiveness, and automation configuration, and should be studied and tooled as one. Together, these findings identify specific points where mitigation remains incomplete. First, for some CVEs, NVD disclosure precedes the fixed release, leaving downstream projects without an adoptable release for weeks to months. Second, when a fixed release precedes disclosure, downstream projects usually miss this window, because adoption follows the routine update process rather than vulnerability disclosure, and most of the delay occurs before a PR or issue is opened. Third, dependabot version updates, which open a PR once a new version is released, are associated with the shortest release-to-adoption delay, yet few projects use them. In future, we will investigate dependency management tools beyond dependabot, decompose exposure into time before a fixed release exists, time before automation is present, and time before adoption, and estimate exposure for pairs without observed adoption dates.

\bibliographystyle{ACM-Reference-Format}
\bibliography{references}


\end{document}